\documentclass[aps,prd,twocolumn,showpacs,amsmath,amssymb]{revtex4}
\usepackage{graphicx}
\usepackage{amsfonts,bm}
\usepackage[english]{babel}
\usepackage{color}
\def\be{\begin{equation}}
\def\te{\end{equation}}
\def\ee{\end{equation}}
\def\ba{\begin{eqnarray}}
\def\bea{\begin{eqnarray}}

\def\tea{\end{eqnarray}}
\def\ea{\end{eqnarray}}
\def\eea{\end{eqnarray}}

\begin{document}

\title{Heavy quark collisional energy loss in the quark-gluon plasma including finite relaxation time}

\author{Mauro El\'ias}
\email{maueli@cab.cnea.gov.ar}
\affiliation{Departamento de F\'isica, Facultad de Ciencias Exactas y Naturales, Universidad de Buenos Aires, CONICET, Cuidad Universitaria, Buenos Aires 1428, Argentina}
\author{J. Peralta-Ramos}
\author{E. Calzetta}
\affiliation{Departamento de F\'isica, Facultad de Ciencias Exactas y Naturales, Universidad de Buenos Aires and IFIBA, CONICET, Cuidad Universitaria, Buenos Aires 1428, Argentina}


\begin{abstract}
In this paper, we calculate the soft-collisional energy loss of heavy quarks traversing the viscous quark-gluon plasma including the effects of a finite relaxation time $\tau_\pi$ on the energy loss. We find that the collisional energy loss depends appreciably on $\tau_\pi$. In particular, for typical values of the viscosity-to-entropy ratio, we show that the energy loss obtained using $\tau_\pi \neq 0$ can be $\sim$ 10$\%$ larger than the one obtained using $\tau_\pi=0$. Moreover, we find that the energy loss obtained using the kinetic theory expression for $\tau_\pi$ is much larger that the one obtained with the $\tau_\pi$ derived from the Anti de Sitter/Conformal Field Theory correspondence. Our results may be relevant in the modeling of heavy quark evolution through the quark-gluon plasma. 
\end{abstract}
\maketitle

\section{Introduction}
\label{intro}

Achieving a deep understanding of the phenomenon of quark energy loss in the quark-gluon plasma (QGP) is of crucial importance for the correct 
interpretation of data on hadron suppression at the Relativistic Heavy Ion Collider (RHIC) and the Large Hadron Collider (LHC), as well as for gaining insight on the thermalization process of matter created in these experimental facilities \cite{eloss1,eloss2,eloss3,eloss4,pde100,pde200,pde500,pde600,pde700,mullerliquid}.

Quark energy loss can occur due to gluon radiation or (hard or soft) collisions. For low energy heavy quarks, the dominant energy loss mechanism is the collision of the heavy quark with the constituents of the QGP (see, for example, Refs. \cite{pde100,pde200,pde500,pde600,pde700,mullerliquid,rapp1,rapp2,rapp3,zapp}). 
We calculate the collisional energy loss of a heavy quark traversing the QGP including the effects of a finite relaxation time $\tau_\pi$ on the energy loss. To our knowledge, this is the first study of collisional energy loss in the QGP including $\tau_\pi$ (see Ref. \cite{carrcoll} for a related study).  

To compute the energy loss including a finite relaxation time $\tau_\pi$, we use the QGP polarization tensor that is derived from the effective hydrodynamic formalism developed by two of us in \cite{dev,app,linking,tensorpolarizacion}. This model, which is constructed from the Entropy Production Variational Method \cite{epvm}, incorporates the effect of higher order velocity gradients into the hydrodynamic description of the QGP, thus extending the applicability of a macroscopic description to strongly out of equilibrium situations, such as early time dynamics of the plasma or the most peripheral collisions. We have shown that the model is able to reproduce the results from kinetic theory even in highly non-equilibrium regimes \cite{net} (see also Ref. \cite{weib} for a study of the Weibel instability based on this model). 

The paper is organized as follows. In section \ref{en} we provide a brief overview of collisional energy loss in the QGP and describe the polarization tensor as obtained from the effective hydrodynamic model. In section \ref{results} we present and discuss our results, and in section \ref{conc} we conclude.

\section{Collisional energy loss}
\label{en}

We will consider an isotropic, non-expanding QGP and compute the collisional energy loss $dE/dx$ of a charm quark that transverses it ($x$ is the distance traveled by the quark). 

The soft-collisional energy loss of a fast particle transversing the QGP can be calculated by linearizing Wong's equations, see Refs. \cite{carrington2,carrington,libro}. In this work we consider a stable plasma, for which all modes are damped and there are no instabilities. This means that the energy loss is solely due to Landau damping, i.e. $\omega=\textbf{k}.\textbf{v}$, where $\omega$ and $\textbf{k}$ are the frequency and wave vector of the excitation, and $\textbf{v}$ is the quark's velocity. The collisional energy loss is then given by
\begin{eqnarray}
-\frac{dE}{dx}=\frac{C_Fg^2}{v}\int \frac{d^3k}{(2\pi)^3} \bigg[ \frac{\omega \   Im(\epsilon_L(\omega,\textbf{k}))}{\textbf{k}^2|\epsilon_L(\omega,\textbf{k})|^2} + \nonumber \\ \frac{\left( v^2-\frac{\omega^2}{\textbf{k}^2} \right)\ Im(\epsilon_T(\omega,\textbf{k}))}{\omega |\epsilon_T(\omega,\textbf{k})-\frac{\textbf{k}^2}{\omega^2}|^2} \bigg] _{\omega=\textbf{k}.\textbf{v}}
\label{5}
\end{eqnarray}
where $C_F$ is the quark constant, and $\epsilon_L$ ($\epsilon_T$) is the longitudinal (transverse) part of the dielectric tensor $\epsilon^{ij}$ (latin indices stand for spatial components). 
The dielectric tensor and its components can be written in terms of the polarization tensor using the equations \ref{dielectrico}-\ref{proydielT}.

\begin{eqnarray}
\label{dielectrico}
\epsilon^{ij} & = & \delta^{ij}+\frac{1}{\omega^2}\Gamma^{ij} \\
\epsilon_L & = &  \frac{k_ik_j}{k_\alpha k^\alpha} \epsilon^{ij} \label{proydielL} \\ \epsilon_T & = & \frac{1}{2}[Tr(\epsilon^{ij})-\epsilon_L] \label{proydielT}
\end{eqnarray}

The polarization tensor characterizes the linear response of the QGP to external perturbations, which in this case is a quark crossing the QGP with constant velocity.
As mentioned in the Introduction, in this paper we shall use the polarization tensor that is derived from the effective hydrodynamic theory developed in Refs. \cite{dev,app,linking,tensorpolarizacion,net}. This theory incorporates the effect of higher order velocity gradients into the hydrodynamic description of the QGP, thus extending its applicability to strongly out of equilibrium regimes (such as early time dynamics, most peripheral collisions and the borders of the fireball). 

The polarization tensor relates the current induced by a small change in the vector potential to the change itself
\begin{equation}
\label{Gpol}
\delta J_a^\mu = - \Gamma^{\mu \nu}_{ab}A_{\nu}^{b} 
\end{equation}
Two of us have shown that in the effective theory developed in \cite{tensorpolarizacion}, the polarization tensor reads
\begin{eqnarray}
\label{Tpol}
\Gamma^{\mu \nu}_{ab} & = & -\delta_{ab}  \frac{\omega^2_{pl}}{(1+W_2W_4)(k^\alpha u_\alpha)^2}[(k^\alpha u_\alpha k^\mu - k^\alpha k_\alpha u^\mu) u^\nu \nonumber \\ &+& k^\alpha
u_\alpha (u^\mu k^\nu - k^\alpha u_\alpha g^{\mu\nu})  - (W_1 +  W_3) (k^\alpha u_\alpha k^\mu \nonumber \\ &-& k^\alpha k_\alpha u^\mu)(k^\alpha u_\alpha k^\nu - k^\alpha k_\alpha u^\nu) ]
\end{eqnarray}
with $W_i$ ($i=1,2,3,4$) given by

\begin{eqnarray}
W_1 = - [ (k_\alpha k^\alpha)^2+(c_s^{-2} - 1)(k^\alpha u_\alpha)^2 ]^{-1} \label{W1} \\
W_2 = \frac{\eta}{(1+c_s^2)\bar{\rho}(k^\alpha u_\alpha)[1+i \tau_\pi(k^\alpha u_\alpha)]} \label{W2222} \\
W_3 = - \frac{W_2(1+4W_1W_4)}{3+3c_s^2W_4/(k^\alpha u_\alpha)^2+4W_2W_4} \label{W3} \\
W_4 = k^\alpha k_\alpha - (k^\alpha u_\alpha)^2 \label{W4}
\end{eqnarray}
%For kinetic theory obtain $\tau_\pi=6\eta/(sT)$ and for AdS/CFT para $N=4\ SYM$ $\tau_\pi=(2-\ln 2)/(2\pi T)$ \cite{roman}.
Here, $k^\mu = (\omega,k_x,k_y,k_z) \equiv (\omega, \textbf{k})$ is the four-wave vector,  $u^\mu= (\sqrt{1+u_x^2+u_y^2+u_z^2}, u_x, u_y, u_z)$ is the plasma four-velocity ($u^\alpha u_\alpha=1$), and $\omega_{pl}^2=\frac{1}{3}m_D^2$ is the plasma frequency with Debye mass $m_D$. $\bar{\rho}$ is the energy density in the (homogeneous) unperturbed plasma. As shown in Ref. \cite{cspaper}, to reproduce the longitudinal modes obtained from a first order Hard-Thermal Loop expansion, the square of the speed of sound $c_s^2$ must be taken as

\begin{equation}
\label{cseff}
c_s^2=\frac{1}{3} \left[ 1+ \frac{1}{2y} \ln \left( \frac{1-y}{1+y} \right) \right]^{-1} + \frac{1}{y^2}
\end{equation}
with $y=\sqrt{\textbf{k}^2}/\omega$, instead of the ideal value $c_s^2 = 1/3$.

We note that the result for $\Gamma^{\mu \nu}$ is the same as that obtained from first-order hydrodynamics (colorless Navier-Stokes) that was obtained in Ref. \cite{ns,priv}, but with an effective shear viscosity $\eta_{eff}$ given by 
\begin{equation}
\label{etaeff}
\eta_{eff}=\frac{\eta}{1+i\tau_\pi (k^\alpha u_\alpha)}
\end{equation}
The appearance of $\eta_{eff}$ in place of $\eta$ is quite natural since $\tau_\pi$ is precisely the relaxation time of the shear tensor $\Pi^{\mu \nu}$ towards its Navier-Stokes value \cite{roman}.
In the context of collisional energy loss, a finite value of $\tau_\pi$ will imply that if a color excitation is produced in the QGP by the passage of a quark, it will decay slower than if $\tau_\pi$ was zero. As we shall see in the next section, this feature has significant effects on the collisional energy loss of quarks crossing the viscous QGP.

\section{Results}
\label{results}

The two most widely used models for the relaxation time $\tau_\pi$ as an input in hydrodynamic simulations of the QGP are the one derived from Boltzmann's equation and the one obtained from the AdS/CFT correspondence for a strongly-coupled $\cal{N}$=4 supersymmetric Yang-Mills plasma. For an in-depth discussion of the relaxation time in weakly and strongly-coupled plasmas, we refer the reader to Refs. \cite{j0,j1,j2,j3}. 

To better understand the impact of a finite relaxation time on quark collisional energy loss, we show results for the energy loss obtained in three cases: $\tau_\pi = 0$, $\tau_\pi|_{\textrm{Boltz}}=5\eta/(sT)$ \cite{j0} and $\tau_\pi|_{\textrm{AdS/CFT}}=(2-\ln 2)/(2\pi T)$ \cite{roman}. Additionally, as a baseline we have calculated the energy loss in the ideal case, in which $\eta/s = 0$ and $\tau_\pi = 0$.

In what follows, unless otherwise stated we consider a charm quark ($m_c = 1.27$ GeV), the plasma at rest $u^\mu=(1,0,0,0)$, and fix the temperature to a typical value of $T=0.3$ GeV and the coupling constant to $g=0.2$.

\begin{figure}[htb]
	\centering
		\includegraphics[trim = 50mm 85mm 40mm 85mm, clip, totalheight=0.31\textheight]{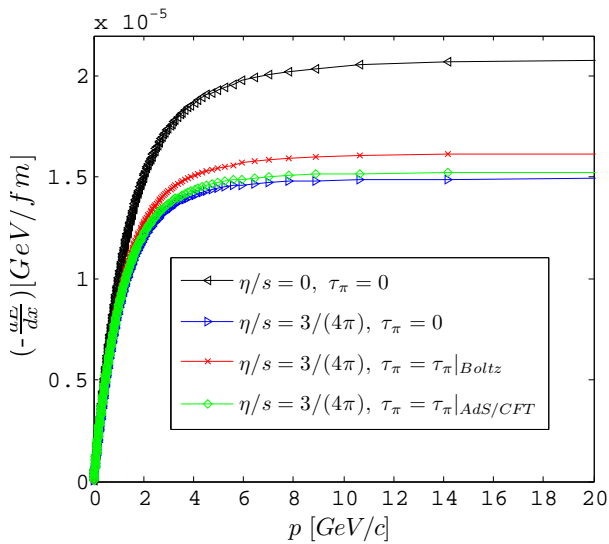}
	\caption{(Color online) Energy loss as a function of quark's momentum for ideal and viscous QGP with $\eta/s = 3/4\pi$, with $\tau_\pi=0$ and $\tau_\pi\neq 0$. The temperature of the plasma is $T=0.3$ GeV.}
	\label{taupis2}
\end{figure}
Figure \ref{taupis2} shows the quark's energy loss as a function of momentum, for ideal and viscous QGP with $\eta/s = 3/4\pi$. It can be observed that the energy loss is maximum for vanishing viscosity; the reason for this behavior will be explained later on when presenting our results with varying values of $\eta/s$. Comparing the ideal and viscous cases, it is seen that the energy loss in the ideal fluid case can be roughly 25$\%$ larger than the one obtained in the viscous case. Our results are consistent with those of Ref. \cite{priv}. This shows that the effects of including the viscosity of the medium on the collisional energy loss of fast particles is significant. 

From Figure \ref{taupis2}, it can also be seen that the effect of $\tau_\pi$ is to increase the energy loss with respect to the $\tau_\pi=0$ case. In the case of the relaxation time corresponding to AdS/CFT, $\tau_\pi|_{\textrm{AdS/CFT}}$, the effect of the relaxation time on the energy loss is very small. In contrast, for $\tau_\pi|_{\textrm{Boltz}}$ the effect of the relaxation time on the energy loss is appreciable. It is seen that for a typical 5 fm medium, the difference in the energy loss for different relaxation times can be at most 10 $\%$.

The difference between the results for $dE/dx$ obtained using both models for $\tau_\pi$ arise because $\tau_\pi|_{\textrm{Boltz}}$ has an explicit dependence on $\eta/s$ and with temperature $T$, whereas $\tau_\pi|_{\textrm{AdS/CFT}}$ only depends on $T$. As mentioned above, in this work the temperature is fixed, so $\tau_\pi|_{\textrm{AdS/CFT}}$ is a constant, but $\tau_\pi|_{\textrm{Boltz}}$ increases with increasing $\eta/s$ and therefore the $\eta_{eff}$ corresponding to kinetic theory becomes smaller than the $\eta_{eff}$ corresponding to AdS/CFT. As a consequence of this, the energy loss obtained using $\tau_\pi|_{\textrm{AdS/CFT}}$ is smaller than the energy loss obtained using $\tau_\pi|_{\textrm{Boltz}}$.

\begin{figure}[htb]
	\centering
		\includegraphics[trim = 45mm 85mm 45mm 85mm, clip, totalheight=0.31\textheight]{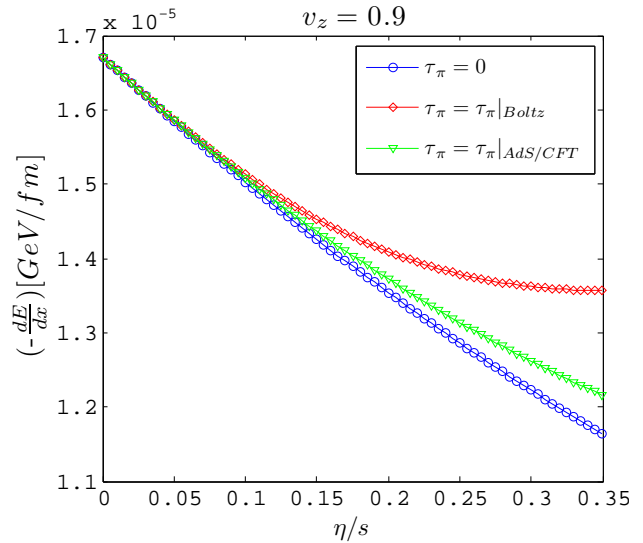}
	\caption{(Color online) Energy loss as a function of $\eta/s$ for a charm quark moving at $v=0.9c$, for the cases with vanishing or finite $\tau_\pi$. The temperature of the plasma is $T=0.3$ GeV.}
	\label{etas}
\end{figure}

Figure \ref{etas} shows the energy loss as a fuction of $\eta/s$, obtained by including or not the time relaxation. The figure corresponds to a quark moving at $v=0.9c$.

Again one can observe that as the value of $\eta/s$ increases, the energy loss decreases. This agrees with the results obtained very recently by Jiang et al \cite{priv}.
In kinetic theory, the viscosity  is $\eta=\bar{p}/(3\sigma_{tr})$ \cite{mullerliquid} where $\bar{p}$ is the mean value of particle momentum in the medium and $\sigma_{tr}$ is transport cross section.
Since the temperature is fixed, so is $\bar{p}$, and therefore when the viscosity increases the cross section decreases, so that the number of collisions with QGP particles decreases, implying less energy loss.

It can be seen from Figure \ref{etas} that the differences in $dE/dx$ between the cases with $\tau_\pi=0$ and $\tau_\pi|_{\textrm{AdS/CFT}}$ are rather small throughout the whole range of values for $\eta/s$ that we consider. The situation is different for the case including $\tau_\pi|_{\textrm{Boltz}}$, for which the differences with the $\tau_\pi=0$ case are significant. For $\eta/s<0.35$, the energy loss calculated by including $\tau_\pi|_{\textrm{Boltz}}$ can be up to 20 $\%$ larger than the corresponding to $\tau_\pi=0$, with the difference between both cases rising with increasing values of $\eta/s$. We note that for the range of typical values for $\eta/s$ at RHIC and LHC, namely $0.08<\eta/s<0.24$, the impact of $\tau_\pi|_{\textrm{Boltz}}$ on $dE/dx$ is at most of 10 $\%$.

\begin{figure}[htb]
	\centering
		\includegraphics[trim = 60mm 85mm 45mm 90mm, clip, totalheight=0.32\textheight]{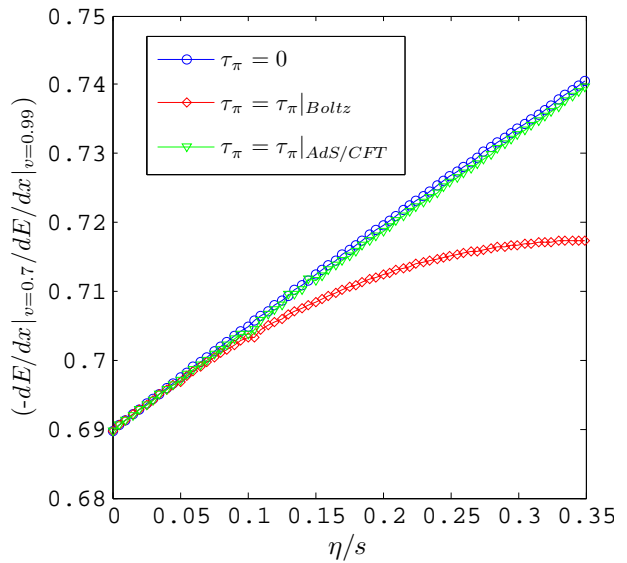}
	\caption{(Color online) Ratio of the energy loss calculated for $v=0.7c$ to that calculated for $v=0.99c$, as a function of $\eta/s$. The temperature of the plasma is $T=0.3$ GeV.}
	\label{coc}
\end{figure}
To better quantify the effect of changing the value of quark velocity $v$ has on collisional energy loss, Figure \ref{coc} shows the ratio of $dE/dx$ calculated for $v=0.7c$ to $dE/dx$ calculated for $v=0.99c$, as a function of $\eta/s$. It is seen that, as expected, a slower quark looses less energy due to collisions than a faster one. The ratio is $~0.7$, and slightly increases with increasing $\eta/s$.

\begin{figure}[htb]
	\centering
		\includegraphics[trim = 45mm 85mm 45mm 85mm, clip, totalheight=0.31\textheight]{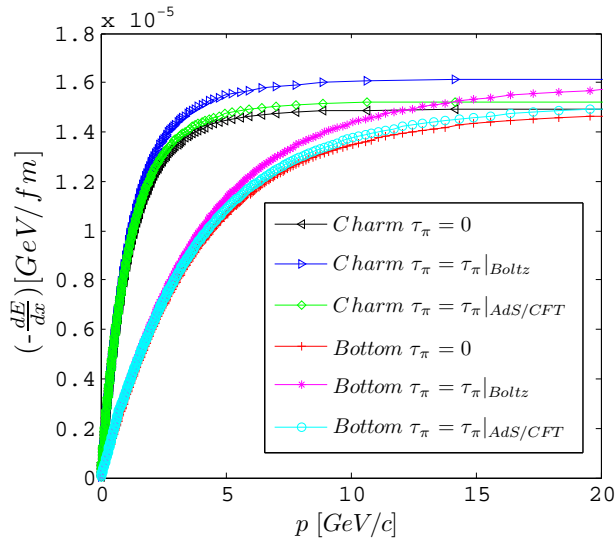}
	\caption{(Color online) Energy loss as a function of momenta for a charm and a bottom quark moving at $v=0.9c$, for the cases with vanishing or finite $\tau_\pi$ with $\eta/s=3/(4\pi)$. The temperature of the plasma is $T=0.3$ GeV.}
	\label{masas}
\end{figure}

To end up, it is interesting to compare the collisional energy loss of a charm quark to that of a bottom quark. Figure \ref{masas} shows $dE/dx$ as a function of momentum for both quarks, obtained with $\tau_\pi=0$, $\tau_\pi|_{\textrm{Boltz}}$ or $\tau_\pi|_{\textrm{AdS/CFT}}$, with $\eta/s=3/(4\pi)$. As expected, the energy loss is larger for the charm quark. We note that the dependence of $dE/dx$ on relaxation time is similar in both cases.

\section{Conclusions}
\label{conc}

We have shown that viscosity effects on the collisional energy loss of heavy quarks through the QGP are important. Comparing the ideal and viscous cases for realistic values of $\eta/s$, it is seen that the energy loss in the ideal fluid case can be roughly 25$\%$ larger than the one obtained in the viscous case.

We have also studied the effect of a finite relaxation time $\tau_\pi$ on collisional energy loss, and found that the effect is appreciable. In particular, we have compared the two most widely used models for $\tau_{\pi}$, namely the one derived from kinetic theory and the one derived from the AdS/CFT correspondence, finding that the effect on energy loss is largest in the former. For realistic values of the shear viscosity to entropy ratio,  the difference in energy loss obtained from both models for $\tau_\pi$ is roughly 10 $\%$. Most importantly, there is a qualitative difference in the way energy loss depends on $\eta/s$ and quark momentum in one and the other model for  $\tau_{\pi}$. This opens up the possibility of using energy deposition to discriminate which model best represents the physics of the quark gluon plasma.

Last but not least, the effect of $\tau_{\pi}$ on the energy deposition on the plasma has a corresponding effect on the back reaction of the plasma on the traversing quark, and therefore modifies the diffusive propagation of the quark itself \cite{qp1,qp2,qp3}. This effect could show up in the analysis of such observables as quarkonium suppression \cite{qs1,qs2}. Of course, a non zero $\tau_{\pi}$ will more generally affect the expansion of the QGP as a whole and therefore its cooling, an effect that also should be taken into account.

\begin{acknowledgements}
We thank Jorge Noronha for useful comments. We also thank Bing-feng Jiang, De-fu Hou and Jia-rong Li for sharing with us their recent work on energy loss in the QGP including viscous corrections.  
This work has been supported in part by ANPCyT, CONICET and UBA under Project UBACYT X032 (Argentina).
 \end{acknowledgements}


\begin{thebibliography}{99}
\bibitem{eloss1} A. Adare et al. (PHENIX Collaboration), Phys. Rev. C \textbf{84}, 054912 (2011).
\bibitem{eloss2} M. J. Tannenbaum, arXiv:1404.6232.



\bibitem{eloss3} P. Gossiaux, J. Aichelin, and T. Gousset, arXiv:1201.4038.

\bibitem{eloss4} S. K. Das, J.-E Alam, and P. Mohanty, Phys. Rev. C \textbf{82}, 014908 (2010).

	\bibitem{pde100}
Markus H. Thoma, J. Phys. G \textbf{26}, 1507 (2000).
%http://arxiv.org/abs/hep-ph/0003016
	\bibitem{pde200}
S. Cao, G. Y. Qin, and S. A. Bass, Phys. Rev. C \textbf{88}, 044907 (2013).
%http://arxiv.org/pdf/1308.0617v3.pdf
	
	\bibitem{pde500}
P. Chakraborty, M. G. Mustafa, and M. H. Thoma, Phys.Rev. C \textbf{75}, 064908 (2007).
	\bibitem{pde600}
M. Djordjevic, Phys.Rev. C \textbf{74}, 064907 (2006).

	\bibitem{pde700}
R. Thomas, B. Kampfer, and G. Soff, Acta Phys. Hung. A \textbf{22}, 83 (2005).
%arXiv:hep-ph/0405189
\bibitem{mullerliquid}
B. Muller, Acta Phys. Polon. B \textbf{38}, 3705 (2007).	

\bibitem{rapp1} H. van Hees and R. Rapp, Phys Rev. C {\bfseries 71}, 034907 (2005). 
\bibitem{rapp2} H. van Hees, V. Greco and R. Rapp, Phys.Rev. C {\bfseries 73}, 034913 (2006).
\bibitem{rapp3} H. van Hees et al., Phys. Rev. Lett. {\bfseries 100}, 192301  (2008).
\bibitem{zapp} K. Zapp, G. Ingelman, J. Rathsman, J. Stachel, Phys. Lett. B {\bfseries 637}, 179 (2006). 
	
	
\bibitem{carrcoll} M.E. Carrington, T. Fugleberg, D. Pickering, M.H. Thoma, Can. J. Phys. \textbf{82}, 671 (2004).
\bibitem{dev} J. Peralta-Ramos and E. Calzetta, Phys. Rev. D {\bfseries 80}, 126002 (2009).
\bibitem{app} J. Peralta-Ramos and E. Calzetta, Phys. Rev. C {\bf 82}, 054905 (2010).
\bibitem{linking} E. Calzetta and J. Peralta-Ramos, Phys. Rev. D {\bfseries 82}, 106003 (2010).
\bibitem{tensorpolarizacion}
	J. Peralta Ramos and E. Calzetta, Phys. Rev. D \textbf{86}, 125024 (2012).
	
\bibitem{epvm} L. M. Martyushev and V. D. Seleznev, Phys. Rep. {\bfseries 426}, 1 (2006).
\bibitem{net} J. Peralta-Ramos and E. Calzetta, Phys. Rev. D {\bfseries 87}, 034003 (2013).
\bibitem{weib} E. Calzetta and J. Peralta-Ramos, Phys. Rev. D {\bfseries 88}, 095010 (2013).

\bibitem{carrington2}	
	M.E. Carrington, K. Deja, and St. Mrowczynski, Acta Phys. Polon. B Proceedings Supplement \textbf{5}, 947 (2012).
\bibitem{carrington}
	M.E. Carrington, K. Deja, and St. Mrowczynski, Acta Phys. Polon. B, Proceedings Supplement \textbf{5}, 343 (2012).
\bibitem{libro} E. Calzetta and B.-L. Hu, {\it Nonequilibrium Quantum Field Theory} (Cambridge University Press, Cambridge, 2008).
\bibitem{mrw}
St. Mrowczynski and M. H. Thoma, Ann. Rev. Nucl. Part. Sci \textbf{57}, 61 (2007).
\bibitem{cspaper}
M. Mannarelli and C. Manuel, Phys. Rev. D \textbf{77}, 054018 (2008).


\bibitem{ns} B.-F. Jiang and J.-R. Li, Nucl. Phys. A {\bfseries 847}, 268 (2010).
\bibitem{priv} B.-F. Jiang, D.-F. Hou, and J.-R Li, arXiv:1405.0083.


\bibitem{roman}
P. Romatschke, Int. J. Mod. Phys. E \textbf{19}, 1 (2010).

\bibitem{j0} G. S. Denicol, T. Koide, and D. H. Rischke, Phys. Rev. Lett. \textbf{105}, 162501 (2010). 
\bibitem{j1} G. S. Denicol, J. Noronha, H. Niemi, and D. H. Rischke, Phys. Rev. D \textbf{83}, 074019 (2011).
\bibitem{j2} J. Noronha and G. S. Denicol, arXiv:1104.2415.

\bibitem{j3} G. Denicol, H. Niemi, J. Noronha, and D. H. Rischke, in {\it Exploring Fundamental Issues in Nuclear Physics: Nuclear Clusters -- Superheavy, Superneutronic, Superstrange, of Anti-Matter}, edited by D. Bandyopadhyay (World Scientific Publishing, 2012).

\bibitem{qp1} C. Young, B. Schenke, S. Jeon and C. Gale, Phys. Rev. C 86, 034905(2012)

\bibitem{qp2} (The JET Collaboration) K. M. Burke, A. Buzzatti, N. Chang, Ch. Gale, M. Gyulassy, U. Heinz, S. Jeon, A. Majumder, B. Muller, G-Y. Qin, B. Schenke, Ch. Shen, X-N. Wang, J. Xu, C. Young and H. Zhang, ArXiv:1312.5003

\bibitem{qp3} S. Cao, G-Y. Qin, S. A. Bass, Phys. Rev. C 88, 044907 (2013).

\bibitem{qs1} B. K. Patra, V. Agotiya and V. Chandra, Eur.Phys. J. C 67, 465 (2010).

\bibitem{qs2} A. Mocsy, P. Petreczky and M. Strickland, Int. J. of Mod. Phys. A, Vol. 28, 1340012 (2013).


\end{thebibliography}
\end{document}